\documentclass[trackchanges]{aastex701}                                                                                                                                                                                                                               \usepackage{array} 
\usepackage{multirow}
\usepackage{amsmath}

\begin{document}

\title{Observation of A Solar Like Magnetic Reconnection Event in an AGN Corona with XRISM}

\author[orcid=0009-0009-1448-7626]{Gal Vardi}
\affiliation{Technion, Department of Physics, Haifa 32000, Israel}
\email[show]{galvardi@campus.technion.ac.il}

\author[orcid=0000-0001-9735-4873]{Ehud Behar}
\affiliation{Technion, Department of Physics, Haifa 32000, Israel}
\email{behar@physics.technion.ac.il}

\author[orcid=0000-0001-9911-7038]{Liyi Gu}
\affiliation{SRON Netherlands Institute for Space Research, Leiden, The Netherlands}
\affiliation{University of Leiden, Leiden Observatory, P.O. Box 9513, NL-2300 RA, Leiden, The Netherlands}
\email{L.Gu@sron.nl}

\author[orcid=0000-0001-5540-2822]{Jelle Kaastra}
\affiliation{SRON Netherlands Institute for Space Research, Leiden, The Netherlands}
\affiliation{University of Leiden, Leiden Observatory, P.O. Box 9513, NL-2300 RA, Leiden, The Netherlands}
\email{J.S.Kaastra@sron.nl}

\author[orcid=0000-0002-1094-3147]{Matteo Guainazzi}
\affiliation{European Space Agency (ESA), European Space Research and Technology Centre (ESTEC), Noordwijk, The Netherlands}
\email{Matteo.Guainazzi@esa.int}

\author[orcid=0000-0002-4992-4664]{Missagh Mehdipour}
\affiliation{Space Telescope Science Institute, Baltimore, MD, USA}
\email{missagh@umich.edu}

\author[orcid=0000-0001-5709-7606]{Keigo Fukumura}
\affiliation{James Madison University, Department of Physics and Astronomy, Harrisonburg, VA 22807, USA}
\email{fukumukx@jmu.edu}

\author[orcid=0000-0003-2869-7682]{Jon Miller}
\affiliation{University of Michigan, Department of Astronomy, 500 Church Street, Ann Arbor, Michigan 48109-1042, USA}
\email{jonmm@umich.edu}   

\author[orcid=0000-0002-1615-179X]{Ari Laor}
\affiliation{Technion, Department of Physics, Haifa 32000, Israel}
\email{laor@physics.technion.ac.il}

\author[orcid=0000-0003-0172-0854]{Erin Kara}
\affiliation{Massachusetts Institute of Technology, Kavli Institute for Astrophysics and Space Research, Cambridge, MA 02139, USA}
\email{ekara@space.mit.edu}

\author[orcid=0000-0003-3894-5889]{Megan E. Eckart}
\affiliation{Lawrence Livermore National Laboratory, CA 94550, USA}
\email{eckart2@llnl.gov}

\author[orcid=0000-0003-2161-0361]{Misaki Mizumoto}
\affiliation{University of Teacher Education Fukuoka, Science Research Education Unit, Fukuoka, Japan}
\email{mizumoto-m@fukuoka-edu.ac.jp}

\author[orcid=0009-0001-9034-6261]{Christos Panagiotou}
\affiliation{Massachusetts Institute of Technology, Kavli Institute for Astrophysics and Space Research, Cambridge, MA 02139, USA}
\email{cpanag@mit.edu}

\author[orcid=0000-0001-5493-7585]{Chen Li}
\affiliation{University of Leiden, Leiden Observatory, P.O. Box 9513, NL-2300 RA, Leiden, The Netherlands}
\email{cli@strw.leidenuniv.nl}

\author[orcid=0000-0002-5701-0811]{Ogawa Shoji}
\affiliation{Institute of Space and Astronautical Science (ISAS), Japan Aerospace Exploration Agency (JAXA), Kanagawa 252-5210, Japan}
\affiliation{Tokyo University of Science, Department of Physics Phvsics and Astronomy}
\email{ogawa@kusastro.kyoto-u.ac.jp}

\author[orcid=0000-0002-8177-6905]{Matilde Signorini}
\affiliation{European Space Agency (ESA), European Space Research and Technology Centre (ESTEC), Noordwijk, The Netherlands}
\email{Matilde.Signorini@esa.int}

\author[orcid=0000-0003-1252-8227]{Keqin Zhao}
\affiliation{University of Leiden, Leiden Observatory, P.O. Box 9513, NL-2300 RA, Leiden, The Netherlands}
\email{kzhao@strw.leidenuniv.nl}

\begin{abstract}
The X-ray source in AGN is commonly referred to as the corona by analogy to stellar coronae. 
The similarities between the two suggest that the heating mechanism of AGN coronae is magnetic reconnection- as in cool stars- but this has not yet been directly observed. This work presents the first observational evidence for a magnetic reconnection flare in an AGN corona. 
We report on a flare 
in NGC\,3783, which was observed with XRISM/Xtend and 
XMM-Newton/EPIC-PN exhibiting distinct temporal evolution in soft ($<2.0$\,keV) and hard ($>2.0$\,keV) X-rays. An Ultra-Fast Outflow (UFO) was detected during the event.
The flare features the Neupert effect- a temporal signature of the hard light curve correlating with the time derivative of the soft light curve, which shows that the flare is powered by magnetic reconnection. The Neupert effect is widely observed in the Sun, with Coronal Mass Ejections (CMEs) playing a role analogous to the UFO. We derive an upper limit of $30 \, R_g$ on the height of the magnetic loop from which the flare originates. 
Using the UFO measured properties to characterize the magnetic field, we obtain $B > 1.3 \times 10^4$\,G for the field annihilated during the flare from total energy considerations, and $B \approx 500$\,G  for the momentary magnetic field during reconnection from a dynamical consideration.

\end{abstract}

%%\keywords{\uat{Galaxies}{573} --- \uat{Cosmology}{343} --- \uat{High Energy astrophysics}{739} --- \uat{Interstellar medium}{847} --- \uat{Stellar astronomy}{1583} --- \uat{Solar physics}{1476}}


\section{Introduction} 

The similarities between solar and black hole coronae have been invoked for many years in magnetically driven models for X-rays in accretion sources \citep{Galeev1979, FieldRogers1993}. 
Both AGN (active galactic nuclei) and stellar coronae are orders of magnitude hotter than their counterpart optical source; the stellar photosphere and AGN accretion disk, respectively. Coronae of cool stars can reach temperatures of $\sim$1MK while the underlying photosphere is only a few 1000\,K.
Similarly, AGN coronae can reach $10^9$\,K,
while the accretion disk temperature is on the order of $10^4 - 10^5$\,K.
Both coronae vary dramatically over short timescales \citep{Gonzalez-Martin2012} down to seconds \citep{Aschwanden2005}, while photospheres of cool stars vary only slightly over years \citep{Shapiro2014}, and AGN accretion disks vary weakly over months \citep{Liu2008}. Thus, the heating and the dynamics of these coronae are distinct from the relative steady state of their underlying counterparts and require unique physical explanations beyond standard stellar and accretion physics. 

Unlike the solar corona, which has been imaged directly showing its short timescale dynamics, in AGN coronae observations are limited to light curves and spectra. Reverberation campaigns \citep[e.g.,][]{Kara2025} and microlensing measurements \citep{Morgan2012} have shown that the AGN X-ray source is of size $\sim 10-20 R_g$, where $R_g = GM/c^2$ is the black hole's gravitational radius. Magnetic fields are naturally invoked in AGN accretion disks due to the streams of ionized gas in the accretion flow and the effect of magnetic rotation instability \citep{Balbus1991}. These fields are believed to be enhanced and twisted as they are dragged into the corona and could eject relativistic jets \citep{Blandford2019}. AGN corona adhere to the Güdel–Benz relation, where the ratio between their radio and X-ray luminosity is $L_R/L_X \sim 10^{-5}$ \citep{LaorBehar2008}. This relation is similar to coronally active stars, where it is attributed to magnetic fields affecting the corona \citep{Guedel1993}. Theoretical works have attributed AGN coronal flares and heating to magnetic reconnection \citep{diMatteo1998}, and Magnetic Hydro-Dynamics (MHD) simulations have shown that magnetic field instabilities can explain the very presence of the corona \citep{Miller2000}. Recent simulations show that magnetic reconnection combined with photon-bubble instability can account for the launching of UFOs \citep{Xu2026}. However, observational evidence for magnetic fields in the AGN corona has only been circumstantial to date. If the analogy with the solar corona is comprehensive, then magnetic fields in the corona are not only present but are the driving force behind the coronal heating, and its short-term variability.

In stellar corona, the conversion of magnetic energy to heat is attributed to magnetic reconnection. In a reconnection event magnetic fields of opposite polarity are driven towards each other by magnetic shearing due to the Sun’s differential rotation. The superposition of opposing polarity fields annihilates the field and converts its energy ($B^2/8\pi$ per unit volume) to kinetic energy of charged particles, mostly protons and electrons. Some of these particles are accelerated outward along the magnetic field lines in a CME. Others follow the magnetic field lines down to the loop footprint, depositing their energy and thus heating the photosphere, which evaporates into the corona \citep{Holman2016}.

The hot solar corona is a persistent source of X-rays. A reconnection event results in an X-ray enhancement, i.e., flare. One unique signature of some solar flares is the Neupert Effect \citep{Neupert1968}, which manifests when the hard X-ray light curve ($L_H$) correlates with the time derivative of the soft X-ray light curve ($L_S$) %as

\begin{equation}
L_H \propto dL_S/dt \label{eq:NeupertDerivative}
\end{equation}

\noindent This correlation is also present between the radio light curve and the derivative of the soft X-ray light curve \citep{Gudel2002}. Neupert explained this by ascribing the hard X-rays and radio to non-thermal downward streaming charged particles (primarily electrons) via non-thermal bremsstrahlung and synchrotron, and the soft X-rays to the heated footprint via thermal bremsstrahlung and emission lines. The effect stems from the hard X-rays representing the instantaneous emission by the particles, while the soft X-rays represent their total energy, so that when integrating over the time since reconnection

\begin{equation}
L_S \propto \int L_H dt
\label{eq:NeupertIntegral}
\end{equation}

The Neupert effect was observed in flares from cool stars, beginning with the M-dwarfs AD Leonis \citep{Hawley1995} and UV Ceti \citep{Guedel1996}, and recently the G star UX-Ari \citep{Inoue2026}. If the microphysics of the AGN corona is similar to that of the solar corona, one would expect to find magnetic reconnection and the Neupert effect in AGN flares as well.

\section{Observations and Data Reduction}

The light curves (LCs) in this paper are a result of a 10-day observation campaign in July 2024 targeting the active galaxy NGC 3783. The leading observatory was XRISM with the additional participation of XMM-Newton, Chandra, NuSTAR, Swift, NICER.
%and HST \citep[see][for details]{Mehdipour2025}. 
The present work is the seventh paper resulting from this campaign %in the "Delving into the depths of NGC 3783" series of publications, which provide a comprehensive analysis of the 2024 observation campaign led by XRISM. 
The previous papers are I. \textit{Kinematic and ionization structure of the highly ionized outflows} \citep{Mehdipour2025}, II. \textit{Cross-calibration of X-ray instruments used in the large, multi-mission observational campaign} \citep{Xrism2025}, III. \textit{Birth of an ultrafast outflow during a soft flare} \citep{Gu2025}, IV. \textit{Mapping of the accretion flow with Fe K$\alpha$ emission lines} \citep{Li2026}, V. \textit{Broad-band modeling of ionized outflows} \citep{Zhao2026} and VI. {\it Evidence for a failed wind based on Fe\,XXV absorption line variability.} (C. Li, et al., submitted).

Here, we focus on the X-ray LCs in the energy range of 0.4-13.0\,keV as measured by Xtend, the X-ray CCD imager aboard XRISM. Data reduction and LC extraction was performed according to the procedure described in the XRISM collaboration’s 2025 cross-calibration paper \citep{Xrism2025} and using the HEASOFT software package. The spectra used in Fig. \ref{fig:absorption} were fitted using Xspec \citep{xspec1996}.

We also utilize XMM-Newton LCs from the European Photon Imaging Camera (EPIC) PN camera and the Optical Monitor (OM). The XMM-Newton/OM was operated in Science User Defined mode, acquiring exposures through the V, B, U, UVW1, UVM2, and UVW2 filters. OM images of NGC\,3783 were processed with the Science Analysis System (SAS) omichain pipeline, applying necessary corrections such as the removal of Modulo-8 fixed pattern noise, to generate images suitable for aperture photometry. Source and background regions were defined to measure count rates, with corrections applied for the point spread function (PSF), coincidence losses, and time-dependent sensitivity. 

XMM-Newton/EPIC-PN was operated in small window mode with the thin filter. EPIC event lists were generated using the standard pipelines and the latest calibration files available at the time the data reduction was performed (May 2026) with SAS version 22.1.0 \citep{SAS2004}. Photometry was performed using a circular aperture with a 40" diameter and the background was determined from a nearby source-free region of the same size.
UV photometry was performed using a circular aperture with a 12" diameter, chosen as the optimal size based on instrument calibration. The background was determined from a nearby source-free region of the same size. 

LC extraction in the hard X-ray in all instruments excludes the 6.0-7.2\,keV energy band, as it contains strong Fe-K emission and absorption lines not directly related to the effects studied in this paper.

\section{Results}

\begin{figure*}[h]
\plotone{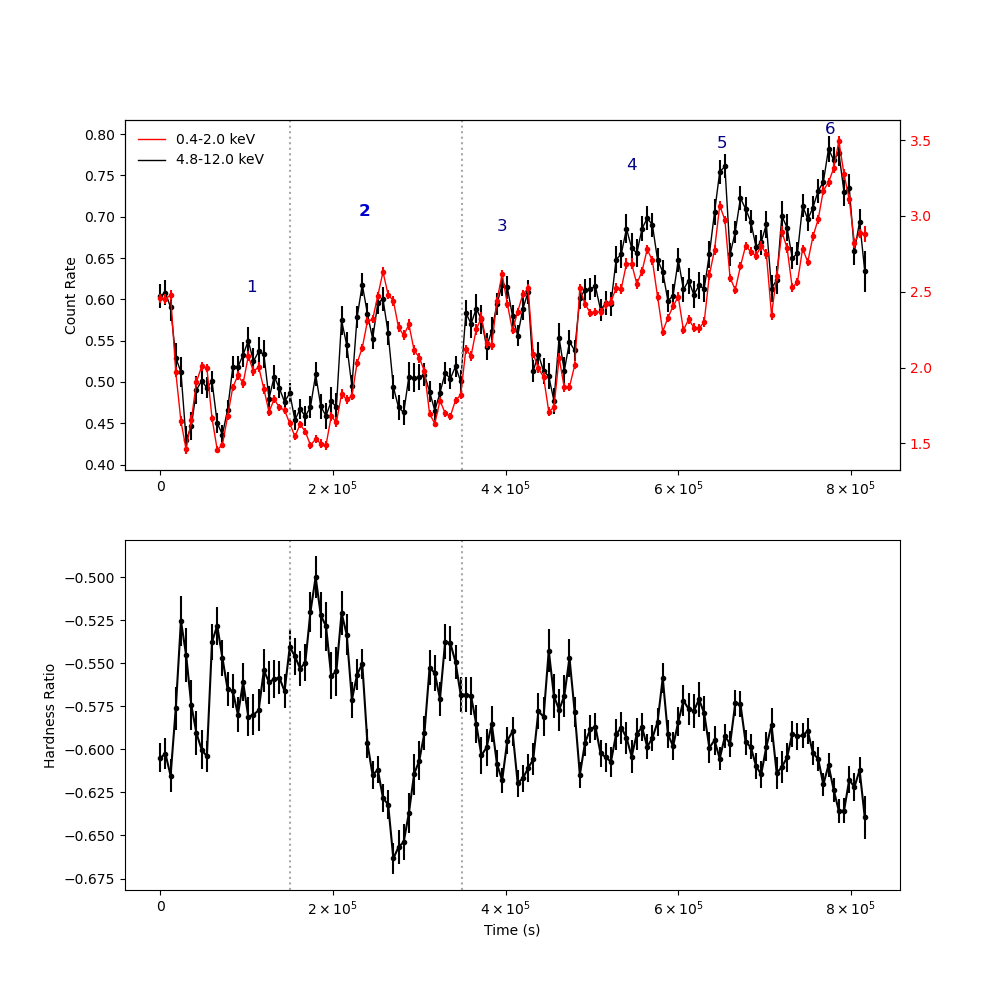}
\caption{(top) XRISM/Xtend light curves (LCs) from the NGC 3783 observation campaign. Hard and soft X-ray LCs are plotted in black and red, respectively. The various peaks in the LC are numbered. The flare discussed in this paper is the highlighted 2nd peak. The segment of the observation shown in Fig.\,\ref{fig:flare_LC} is marked by the dotted lines. Other than during this segment, the hard and soft X-ray LCs are highly correlated through most of the observation (Table \ref{tab:peak_corr}). (bottom) The hardness ratio $(H-S)/(H+S)$. The hardness of the relevant flare is distinctly different from the rest of the observation.}
\label{fig:flares_hardness}
\end{figure*}

\begin{figure*}[h]
\plotone{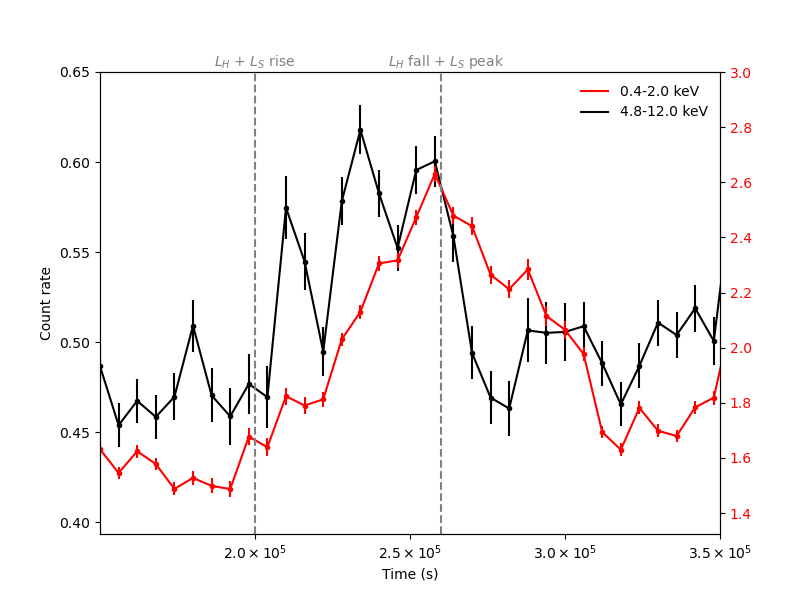}
\caption{XRISM/Xtend soft (red) and hard (black) LCs of NGC 3783 during the flare, binned to 6\,ks time steps. Time shown starts with the XRISM observation. The beginning of the rise in luminosity in both bands at $\sim 2\times 10^5$\,s and the sharp drop in hard X-ray at $\sim 2.6 \times 10^5$ s as defined by \cite{Gu2025} are marked. The rise and fall of the hard X-rays are steeper than those of the soft X-rays, suggesting different emission mechanisms. We interpret the hard X-rays ($L_H$) as originating from non-thermal electrons following a magnetic reconnection event, while soft X-rays ($L_S$) are emitted in the alleged loop footprints. In the hard X-rays the 6.0-7.2 keV range is excluded as it contains strong emission and absorption lines unrelated to the Neupert effect.}
\label{fig:flare_LC}
\end{figure*}

XRISM/Xtend observations of NGC 3783 show multiple X-ray flares with timescales of $\sim$ 1-2 days (Fig. \ref{fig:flares_hardness}), one of which coincides with a UFO \citep{Gu2025}. This particular flare, No.\,2 in Fig.\,\ref{fig:flares_hardness} its LC shown in Fig.\,\ref{fig:flare_LC}, exhibits key differences from the other flares. Unlike the other flares where the hard and soft X-ray LCs are highly correlated, in this flare the hard X-rays as a whole precede the soft X-rays by $\sim$30 ks (Table \ref{tab:peak_corr} in Appendix \ref{appendix_correlations}). 
The hard and soft X-ray peaks are differently shaped,
resulting in a significant decline in hardness ratio compared to the rest of the observation. The increase in hardness leading to the flare and the following softness during the flare's peak and decay are unlike any other hardness behavior during this observation \citep[see Fig.\,2 in][]{Gu2025}.%, and atypical of a passing absorber
After the peak of the flare at $\sim 2.6 \times 10^5$\,s into the observation (Fig.\,\ref{fig:flare_LC}) the hard X-rays drop sharply, followed by a more gradual decay in soft X-rays. The correspondence of the soft X-ray peak with the rapid drop in hard X-rays is consistent with the integral description of the Neupert effect (Eq.\,\ref{eq:NeupertIntegral}). 

In search for the differential Neupert effect (Eq.\,\ref{eq:NeupertDerivative}) we calculate the cross-correlation between the hard X-ray LC and the time derivative of the soft X-ray LC. The results presented in Fig.\,\ref{fig:neupert} show significant correlation with a %correlation 
strength of 0.84 and p-value of 0.002 (see Appendix\,\ref{appendix_ccr}), confirming the Neupert effect, a key signature of magnetic reconnection.

\begin{figure*}[h]
\plotone{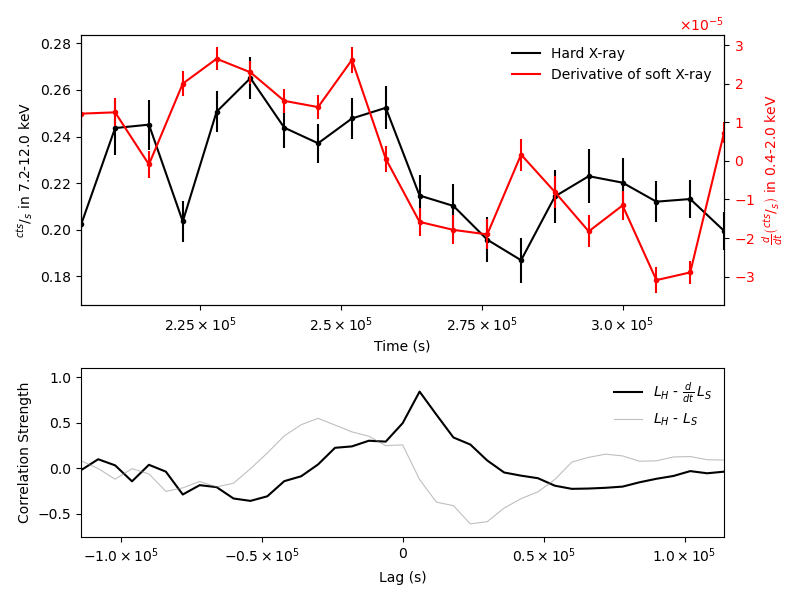}
\caption{Correlation between the hard X-ray LC and the derivative of the soft X-ray LC demonstrating the Neupert effect (Eq.\,\ref{eq:NeupertDerivative}). (top) Black- XRISM/Xtend hard X-ray light curve in the 7.2-12.0 keV band. Red- Time derivative of the XRISM/Xtend soft X-ray light curve in the 0.4-2.0 keV band.  (bottom) Black- Cross correlation between the hard X-ray LC ($L_H$) and the derivative of the soft X-ray LC ($L_S$). Their lag smaller than the 6\,ks bin size. Grey- The insignificant cross correlation between the hard and soft X-ray LCs shown in Fig.\,\ref{fig:flare_LC}.
}
\label{fig:neupert}
\end{figure*}

\begin{figure*}[h]
%\plotone{ratio.png}
\plotone{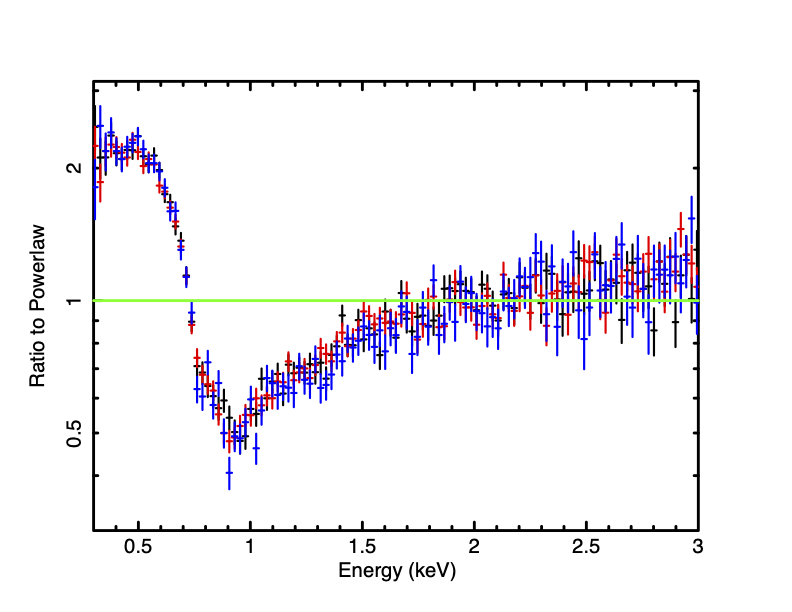}
\caption{XRISM/Xtend spectra of the flare in three time slices; %(c.f.,  Fig.\,\ref{fig:flare_LC})
(black) Hard state before the flare at $(1.2-1.8)\times10^5\,$s, (red) Flare rise at $(2.0-2.6)\times10^5\,$s and (blue) Soft flare decay at $(2.8-3.2)\times10^5\,$s.
The spectra are presented as ratios to a powerlaw, each fitted above 3.0\,keV. Clearly, the deep absorption trough and soft excess below 2\,keV do not change throughout the flare.}
\label{fig:absorption}
\end{figure*}

We rule out other variability effects as the cause for $L_S$ lagging $L_H$ and for $L_H \propto dL_S/dt$. Reflection would have shown a correlation between the soft and hard X-ray LCs, but that is not the case as seen from the faint gray line in the bottom panel of Fig.\,\ref{fig:neupert}, with a peak of 0.55 and p-value of 0.15.
Transient absorption, or typical soft-excess variability are also not observed here.
Fig.\,\ref{fig:absorption} shows that no change in the absorption, nor in the soft-excess occurred during the present flare. The hardness variability is therefore due to the power-law, namely the population of Comptonizing electrons.
Needless to say, none of these alternative variability sources would have resulted in $L_H \propto dL_S/dt$ and a launch of a UFO, which are natural consequences of a magnetic reconnection event.
%\textcolor{orange}{This is totally expected since ionized absorbing outflows are much further away from the center and thus do not vary within hours as observed here, but (specifically in NGC\,3783) on timescales of years \citep{Zhao2026}, months \citep{Mehdipour2017}, or weeks \citep{Markowitz2014}. 
%The present variability is not the typical soft-excess variability of this source either, which has been observed on significantly longer timescales and lower amplitudes \citep{Netzer2003}.
Finally, we also examine the correlations in other flares and in different LC and LC-derivative combinations (Appendix \ref{appendix_correlations}). No other such combination or flare exhibits a significant correlation.

\begin{figure*}[h]
\centering
\plotone{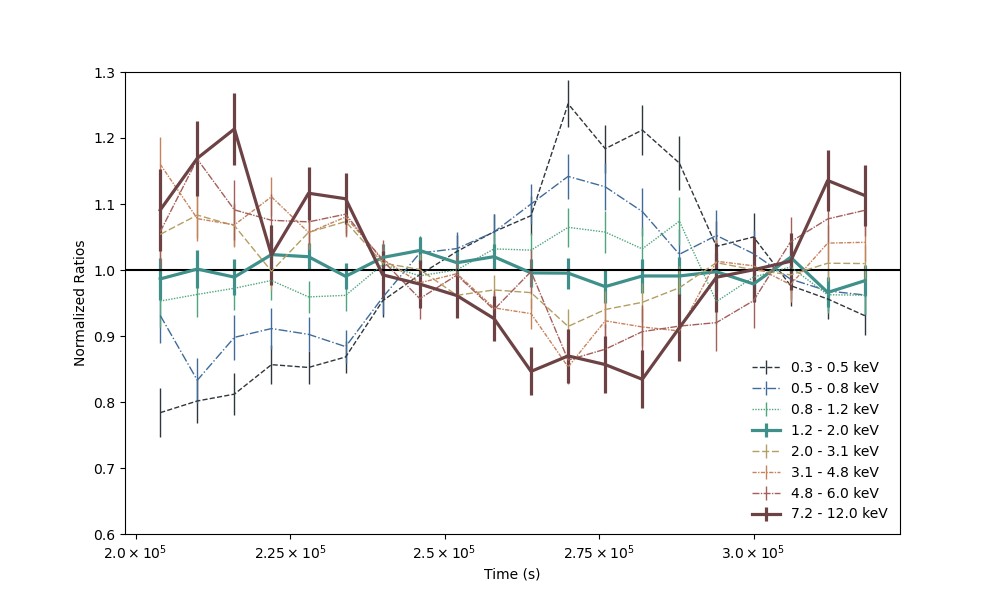}
\caption{Ratio between energy-sliced LCs of the flare in Fig.\,\ref{fig:flare_LC} and the full LC at 0.3-12.0 keV. Slices have similar logarithmic widths. All light curves have been normalized to a mean of 1 for a meaningful ratio.
The different temporal behavior of the hard and soft X-rays can be clearly seen. Highlighted are the bands that yield the most significant correlation between the hard X-ray (7.2-12\,keV) LC and the derivative of the soft X-ray (1.2-2.0\,keV) LC (Eq. 1). The latter band has the highest count rate, and thus is closest to the full LC shape.
There are several common features within the soft ($<$ 2 keV) bands and distinctively within the hard ($>$ 2 keV) bands, indicating the soft and hard X-ray are emitted by different physical mechanisms. The 6.0-7.2\,keV range is not presented as it contains strong Fe-K emission and absorption lines unrelated to the Neupert effect.}
\label{fig:energy_bands}
\end{figure*}

Different choices of soft and hard X-ray bands still present the Neupert effect, however the most significant correlation is found between the 1.2-2.0\,keV and 7.2-12.0\,keV bands. 
Inspection of the light curves at different bands reveals similar temporal behavior for all soft X-ray bands under 2.0\,keV, and a different behavior for all hard X-ray bands above 2.0\,keV, as shown in Fig.\,\ref{fig:energy_bands}. This matches our interpretation of the flare being caused by two separate emission mechanisms along the loop and at the loop footprint, as in the Sun, and allows identification of the characteristic energies emitted in both processes. 
Increasing the soft band to include 0.4-2.0\,keV increases the SNR, and insignificantly increases the $p$ value from 0.0018 to 0.0021, hence the choice of energy bands in Fig. \ref{fig:neupert}.

\begin{figure}[h]
\gridline{\fig{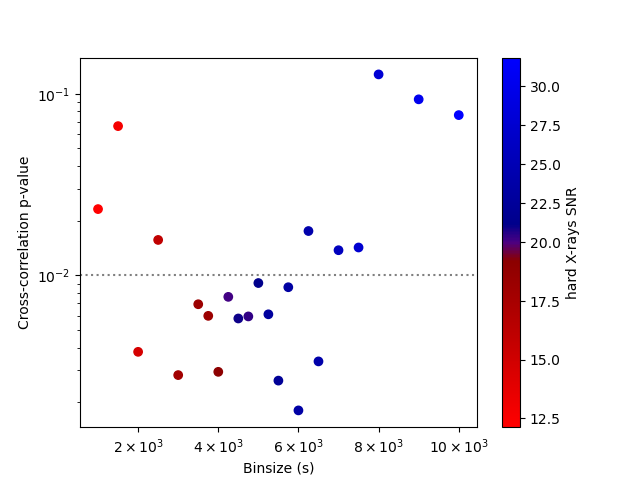}{0.6\textwidth}{}}
\caption{The p-value (vertical axis) of the cross correlation between the hard X-ray (7.2-12.0\,keV) LC and the time derivative of the soft X-ray (0.4-2.0\,keV) LC (Eq.\,\ref{eq:NeupertDerivative}) and the SNR (color scale) of the hard X-ray LC at various temporal binning choices. The strongest correlation is obtained with 6\,ks bins, which is used in all other LCs in this paper. Most choices of bin sizes between 4 and 8\,ks yield a significant correlation with p-value $<$ 0.01 (including binning by the XRISM satellite orbit at 5747 s), and all have p-value $<0.05$. This indicates the observed Neupert effect is not an artifact of binning choice. In bins smaller than 4\,ks the hard X-ray LC SNR drops below 20, making the correlation unreliable.}
\label{fig:binning}
\end{figure}

We then tested different temporal binnings. A significant correlation between the hard X-rays and the derivative of the soft X-rays holds for every temporal binning of 4-8\,ks (Fig. \ref{fig:binning}), proving the Neupert effect is not caused by aliasing or binning choice. When binned finer than 4\,ks, the SNR is too low for a meaningful correlation analysis. Above 8\,ks the features of the flare's light curve and thus the difference between hard and soft X-rays become ambiguous. 
In all of these binning choices, the lag between the hard X-rays and the derivative of the soft X-rays is less than the bin size. Additionally, the time of the beginning of the flare in both hard and soft X-rays
at $\sim 2.0 \times 10^5$\,s is hardly distinguishable (Fig.\,\ref{fig:flare_LC}). These two phenomena indicate that the time difference between both emission processes is smaller than the minimal bin size of 4\,ks. 

Adopting the magnetic reconnection interpretation, the time difference is attributed to the travel time of charged particles along the magnetic loop. With NGC\,3783’s black hole mass of $M = 2.8 \times 10^7 \, M_\odot$ \citep{Bentz2021} its gravitational radius is $R_g = GM/c^2 = 4.1 \times 10^{12}$\,cm, where $G$ is the gravitational constant and $c$ the speed of light. Assuming the particles travel close to the speed of light and taking 4\,ks as a conservative upper limit, the distance between the reconnection event and the loop footprint is thus $ct < 30 \, R_g$. This distance is consistent with MHD simulations of coronal structures in AGN with similar properties \citep{Igarashi2024}.

Since XMM-Newton/EPIC-PN observed the same flare as XRISM, we also check for the Neupert effect in its light curves. Since the count rate above 7.2\,keV in EPIC-PN is too low, for the hard X-ray LC we use instead the 4.8-6\,keV band.
These LCs produce the same correlation between the hard X-ray and soft X-ray derivative as those of XRISM/Xtend with a correlation strength of 0.92 and p-value of 0.02, corroborating the detection of the Neupert effect. We ascribe the somewhat higher p-value to the different hard X-ray bands, see the optimal band selection in  Fig.\,\ref{fig:energy_bands}. 

\begin{figure*}[h]
\centering
\plotone{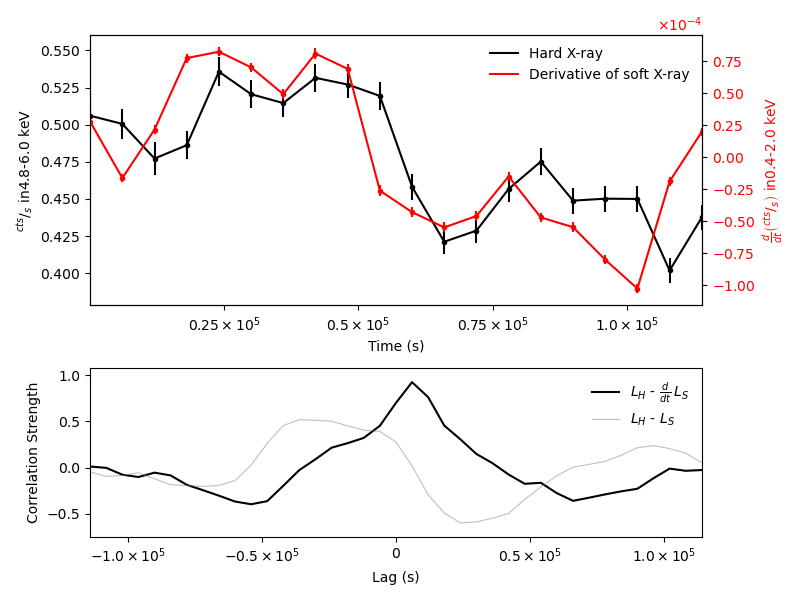}
\caption{Correlation using XMM-Newton EPIC-PN data between hard X-ray LC and the derivative of the soft X-ray LC demonstrating the Neupert effect (Eq.\,\ref{eq:NeupertDerivative}). (top) Black- Hard X-ray LC in the 4.8-6.0\,keV band. Red- Time derivative of the soft X-ray LC in the 0.4-2.0 keV band. Time shown is from the start of the XMM observation. (bottom) Black- Cross correlation between the hard X-ray LC ($L_H$) and the derivative of the soft X-ray LC ($L_S$), with a lag smaller than the 6\,ks bin size. Grey- the insignificant cross correlation between the hard and soft X-ray LCs. This behavior mirrors the XRISM/Xtend LCs.}
\label{fig:neupert_XMM}
\end{figure*}

\begin{figure}[h]
\gridline{\fig{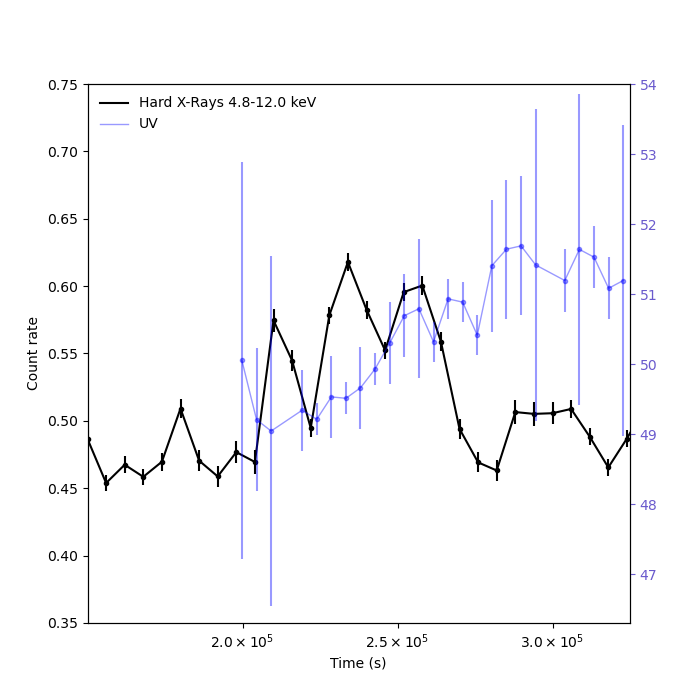}{0.6\textwidth}{}}
\caption{Hard X-ray (black) and UV (violet) LCs during the NGC\,3783 flare observed by XRISM/Xtend and the XMM-Newton Optical Monitor (UVW1, UVW2, and UVM1 filters combined).
The UV flare lags the hard X-ray one by $\sim60$\,ks, although UV uncertainties impede an accurate estimate.
Such a lag implies the UV-emitting region is $\sim400\,R_g$ from the reconnection event.
The gradual rise of the UV flare can represent the size of the UV-emitting region of $\sim180-360\,R_g$.}
\label{fig:UV}
\end{figure}

The UV light curve from XMM’s Optical Monitor resembles a flare lagging after the hard X-ray flare by approximately 60\,ks, but is significantly weaker than the X-ray flare when compared to the quiescent UV flux. While the SNR of the UV impedes a formal correlation, one can  tentatively identify shared features between the UV and hard X-ray LCs that are absent in the soft X-rays LC (see Fig.\,\ref{fig:UV}). 
The similarity in the shape of the two curves is consistent with the UV flare responding to the hard X-ray flare, as the UV-emitting region of the disk is illuminated by the hard X-rays. 
This lag can be understood as the light travel time between the reconnection event in the corona and the UV-emitting region of the disk. 
Such a lag implies the UV-emitting region is $\sim400\,R_g$ from the reconnection event.
Furthermore, the rise of the UV flare is slower than that of the hard X-ray flare due to the large size of the UV-emitting region. If we interpret this as the light travel time across the UV-emitting region of the disk, we can estimate it to be $\sim180-360\,R_g$ in size.
This is well within the range 
of $100-500\,R_g$ quoted in \citet{Kara2025}.

\section{Discussion}

\begin{figure}[h]
\gridline{\fig{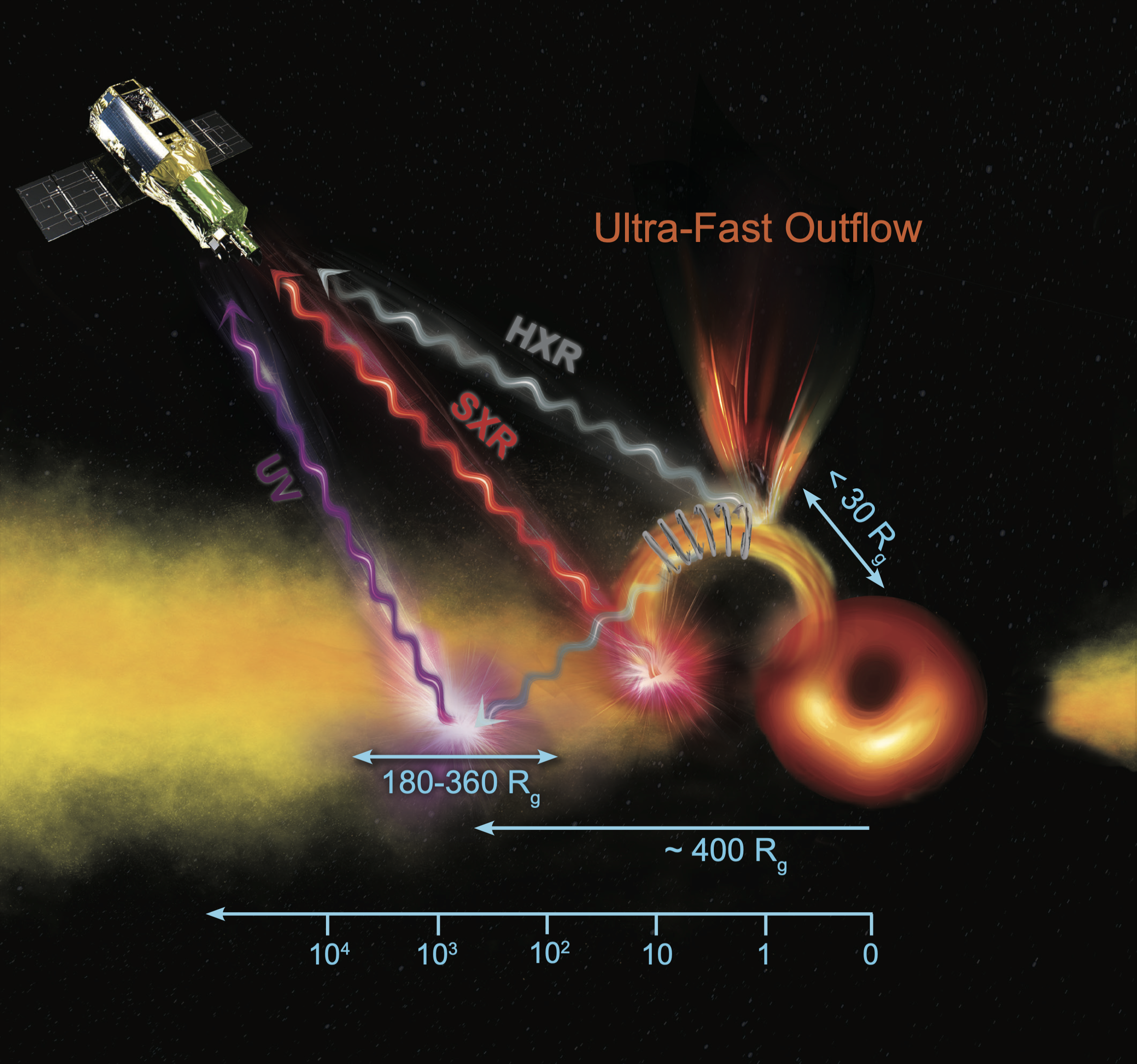}{0.8\textwidth}{}}
\caption{Schematic illustration of the detected magnetic reconnection event and ensuing processes. When the magnetic loop destabilizes, magnetic energy is converted into particle kinetic energy. Some of the material in the loop is ejected as an outflow and some travels along the magnetic field lines emitting hard X-rays. When this material reaches denser regions at the loop footprints, it heats the corona, thus emitting soft X-rays. Some of the hard X-rays illuminate the disk, heating it, and enhancing its UV emission. The scales in this figure are directly derived from the time delays between different phases of the hard X-rays, soft X-rays, and UV LCs. The gravitational radius of NGC\,3783 is $R_g \approx 4.1 \times 10^{13}$ cm. The illustration is presented on a logarithmic scale to account for the difference in scale between corona and disk. (Credit: Dina Mid. the image of the central black hole is courtesy of the EHT collaboration.)}
\label{fig:schematic}
\end{figure}

The observed Neupert effect as a solar-like magnetic reconnection event ties together the flares in X-rays and UV as well as the UFO observed by \cite{Gu2025}. This event launches a CME-like outflow from a magnetic loop $<30 \, R_g$ above the corona, while also accelerating protons and electrons along the loop field lines down to the corona. These particles emit hard X-rays, but eventually reach the loop footprint, emitting soft X-rays through inverse Compton scattering of photons from the heated region or from the disk. Due to their different height scale, some of the hard X-ray radiation illuminates the disk, while the soft X-rays from the loop footprint do not. As a result, only the hard X-rays are reflected in the form of UV from the disk. Fig.\,\ref{fig:schematic} presents a schematic illustration of these processes and length scales.

In the sun the hard X-rays originate in bremsstrahlung from the non-thermal charged particles, 
while the soft X-rays are due to thermal bremsstrahlung and emission lines from the loop footprints. Due to the high temperatures of AGN coronae, and the strong ambient radiation field, here both 
the non-thermal (hard) and thermalized (soft) charged particles emit X-ray photons through inverse Compton. In the sun, the non-thermal electrons travel at sub-relativistic velocities of $\sim0.5\,c$ \citep{Reid2014} with no dependence on flare luminosity or magnetic loop size \citep{Aschwanden1996}. AGN flares are orders of magnitude more luminous, and relativistic beaming may or may not be present. In any event, both hard and soft X-rays are continuum emissions, and are thus differentiated by their distinct temporal, and not spectral, behaviors. In fact, the energy bands exhibiting the Neupert effect most strongly (Fig.\,\ref{fig:energy_bands}) provide the characteristic energies for the two origins. In the present flare, the soft and hard X-rays are differentiated around 2\,keV.

Considering the height of the flare loop $< 30\,R_g$ and estimating the size of the corona at $\sim 10\, R_g$, we conclude that the UFO was launched with the flare at a distance of $r_0 \sim 10-40\,R_g$ from the black hole. 
The outflow was observed $10^5$\,s later with a velocity of $v_{UFO} \sim 0.2c$ \citep{Gu2025}, thus at a distance of $r_{UFO} = r_0 + 0.2c \cdot 10^5 = 160-190\,R_g$ or $(6.6 - 7.8) \times 10^{16}$\,cm from the black hole. 
For this UFO, we adopt an ionization parameter $\xi = L/ (nr^2) \sim 10^3$\,erg\,s$^{-1}$cm, ionizing luminosity $L = 2 \times 10^{44}$\,erg\,s$^{-1}$ and hydrogen column density $N_H = 10^{23}$\,cm$^{-2}$ \citep{Gu2025} to estimate its number density

\begin{equation}
    n = \frac{L}{\xi r_{UFO}^2} \approx 2.2-3.1 \times 10^{11} \, \mathrm{cm}^{-3}
\end{equation}

and thickness

\begin{equation}
    \Delta R = \frac{N_H}{n} = 3.2 \times 10^{11}\,\mathrm{cm.} 
\end{equation}

\noindent Clearly, $\Delta R << r_{UFO}$, indicating the outflow is a geometrically thin blob.

Since only a fraction $\eta = E_{UFO}/E_{mag} < 1$ of the magnetic energy released in the reconnection event is transformed into the outflow's kinetic energy, the energy densities of the outflow and of this magnetic field are related as

\begin{equation}
    \frac{1}{2}n m_p v_{UFO}^2 = \eta \frac{B^2}{8\pi}
\end{equation}

\noindent where $m_p$ is the proton mass. The constraint $\eta<1$ yields a lower limit on the magnetic field strength of $B > v_{UFO} \sqrt{4\pi n m_p} \approx 1.3 \times 10^4$ G. 
Assuming $\eta = 10\%$, %the fraction of the initial magnetic energy in the UFO is 
similar to the sun
%, where it is estimated at $\eta \sim 0.1$ 
\citep{Aschwanden2017}, we can estimate the strength of the magnetic field at $4\times 10^4$ G.

This magnetic field is orders of magnitude higher than the  field expected in radio-quiet AGN coronae \citep[e.g.,][]{delPalacio2025}. Such a high field produces a local energy density that exceeds that of the seed-photon field that inverse Compton-scatters to produce the hard X-rays. If prevalent throughout the corona, this field would make the AGN radio-loud. However, in solar flares the magnetic fields vary significantly in different parts of active regions \citep{Sun2012}, reaching a local maximum near the reconnection site that is orders of magnitude stronger than in the quiescent corona  \citep{Chen2020}. Similarly, the strong magnetic field in the magnetic reconnection event in AGN is highly localized in the looptop. The hard X-rays are due to inverse Compton scattering throughout the loop, and are emitted from a much larger volume.

%In CMEs, acceleration is not attributed to Alfvén waves. 
Alternatively, we can utilize the measured peak acceleration of the UFO $a_{\rm UFO}$ to express a collective equation of motion (force per unit area), due to the magnetic pressure as

\begin{equation}
    m_p a_{\rm UFO} N_{\rm H} = \frac{B^2}{8\pi}
\end{equation}

%were $a_{UFO}$ is the acceleration, N is the number of ions in the UFO and A is its surface area. 
%where acceleration is gained at varying rates 
\noindent The measured peak acceleration measured by \citet{Gu2025} is $6\times 10^4$\,cm\,s$^{-2}$, yielding $B\sim500$\,G, which is considerably lower than the above energy conversion estimate.
In reality, $B$ varies during the reconnection event due to changes in the configuration of the magnetic field \citep{Hu2014}, and this value represents the momentary strength of the magnetic field during reconnection. This, quoting a single $B$ value may be misleading.
%$10^5\,$s after the flare is estimated at $0.6\,km/s^2$. If the acceleration, as in CMEs, is from the magnetic pressure then
%Considering $A/N=N_H$, we estimate the magnetic field $10^5\,$s after the flare at $B\sim500G$.
It remains to be seen whether the magnetic field, magnetic loop size and characteristic energies derived from this observation are specific to this event or characteristic of magnetic reconnection in AGN.

The synchrotron cooling time of relativistic particles in a magnetic field is given by

\begin{equation}
    t = \left( \frac{2 e^4 B^2}{3 m^3 c^5} \gamma_0 \right)^{-1}
\end{equation}

\noindent where $m$ and $\gamma_0$ are the relativistic mass and Lorentz factor of the particles \citep{Rybicki1986}. For $\gamma_0=100, B=10^4$\,G the electrons cooling time is on a scale of seconds, while for protons ($m_p = 1836 \, m_e$) it can be many years. 
Since 2/3 of the energy in magnetic reconnection resides in the protons \citep{Yamada2014}, the electrons are continuously re-energized by the protons, enabling the X-ray flare to continue for $\sim100$\,ks.
Synchrotron cooling therefore does not impede the electrons' travel along the loop.
If the magnetic field and radiation field are in equipartition, then Compton cooling is also negligible. Most of the flare energy is therefore deposited in the corona and in the outflow.

In the solar corona, the Neupert effect is observed in about 50$\%$ of flares \citep{Veronig2002}. The reason for this rate of occurrence 
remains an open question. In AGN flares, it still needs to be determined. We suggest several physical effects that may limit detectability. For one, the hard X-rays emitted by downward streaming electrons are strongly beamed \citep{Brown1972}. Second, reconnection in small loops produces short-lived hard flares in which the Neupert effect cannot be temporally resolved. In other flares, the differentiating energy between the emission of downstreaming particles and that of the loop footprints (in the present flare around 2 keV) may be outside the sensitivity energy range of the instrument. Observation of a solar flare with a failed CME by \citet{Gou2026} showed that simultaneous reconnection processes can interfere with one another, leading to smaller magnetic loops. In such events, the charged particles do not cleanly follow the magnetic loop structure, which may interfere with the Neupert effect. Such flares can result in strong X-ray enhancements, which may explain some strong AGN flares that do not exhibit the Neupert effect or UFOs. 

\section{Conclusions}
While the existence of magnetic reconnection in AGN corona has been long hypothesized, this paper presents its first direct detection through the Neupert effect.
Our findings can be summarized as follows:
\begin{itemize}
    \item We report the detection of the Neupert effect, namely $dL_S/dt \propto L_H$ in a flare observed in NGC\,3783 with the XRISM and XMM-Newtong telescopes.
    \item The measured lag between the two bands indicates that a magnetic reconnection event occurred, in which magnetic energy was converted into the kinetic energy of fast charged particles $<30\,R_g$ above the magnetic loop footprints. 
    \item The simultaneous launching of a UFO by the flare associates it with stellar-like CMEs.
    \item The inferred energy of the UFO is used to estimate the magnetic field at the site of reconnection to be $B \sim 10^4$\,G, while its inferred acceleration during the flare suggests a lower momentary value of $B = 500$\,G. In reality, we expect a gradient of magnetic fields in the corona ranging over several orders of magnitude in $B$.
\end{itemize}

The present discovery of the Neupert effect suggests magnetic reconnection can explain both the AGN coronal heating mechanism, and UFOs as CMEs. It opens new paths to investigating the AGN corona through characterization of additional Neupert flares and their occurrence rate, as well as their implications for otherwise difficult to access coronal properties such as length scales and magnetic fields.

\begin{acknowledgments}
This research was supported by The Israel Science Foundation grant No. 2617/25.
SRON is supported financially by NWO, the Netherlands Organization for Scientific Research.
Part of this work was performed under the auspices of the U.S. Department of Energy by Lawrence Livermore National Laboratory under Contract DE-AC52-07NA27344.
M.M. is supported by JSPS KAKENHI JP26K17197 and the Yamada Science Foundation.
\end{acknowledgments}

\appendix
\section{Uniqueness of the Neupert Effect in one Flare}
\label{appendix_correlations}

To substantiate the presence of the Neupert effect, and in one flare only,
we compare in Table \ref{tab:peak_corr} all of the flares in the Xrism/Xtend observation of NGC\,3783 (Fig.\,\ref{fig:flares_hardness}). 
It can be seen that Flare \#2 (Fig.\,\ref{fig:flare_LC}) shows the weakest correlation between the hard and soft X-rays. Moreover, it is the only flare in which the soft X-rays lag the hard X-rays. All other flares are consistent with both bands flaring at exactly the same time.  Most importantly, no other flare shows as strong a correlation between the hard X-ray LC and the {\it derivative} of the soft X-ray LC.

\begin{deluxetable*}{ccccccccc}
\label{tab:peak_corr}
\tablecaption{Correlations between the hard X-ray LC ($L_H$) and the soft X-ray LC ($L_S$) and its derivative around each of the six peaks (Fig \ref{fig:flares_hardness}) during the XRISM observation of NGC\,3783. The correlations are calculated using XRISM/Xtend light curves at 6\,ks binning. The hard and soft X-ray bands are 7.2-12.0\,keV and 0.4-2.0\,keV, respectively, as these are the bands maximizing correlation significance and SNR (see Fig. \ref{fig:energy_bands}). 
Flare \#2, which features the Neupert effect and is highlighted, stands out in several ways (see text for details).
}
\tablehead{
  \colhead{Peak} & \colhead{Start time} & \colhead{End time} & 
  \colhead{$L_H$-$L_S$} & \colhead{$L_H$-$L_S$} & \colhead{$L_H$-$L_S$} & 
  \colhead{$L_H$-$\frac{d}{dt}L_S$} & \colhead{$L_H$-$\frac{d}{dt}L_S$} \\[-8pt]
  \colhead{} & \colhead{(ks)} & \colhead{(ks)} & 
  \colhead{corr. strength} & \colhead{p-value} & \colhead{lag (ks)} & 
  \colhead{corr. strength} & \colhead{p-value}
}
\startdata
1 & 55  & 190 & 0.64 & 0.005 & 0   & 0.40 & 0.1   \\
\textbf{2} & \textbf{200} & \textbf{320} & \textbf{0.55} & \textbf{0.15} & \textbf{-30} & \textbf{0.83} & \textbf{0.002} \\
3 & 320 & 460 & 0.77 & 0.009 & 0   & 0.45 & 0.06  \\
4 & 360 & 590 & 0.83 & 0.03  & 0   & 0.31 & 0.3   \\
5 & 620 & 670 & 0.94 & 0.05  & 0   & 0.52 & 0.4   \\
6 & 720 & 820 & 0.80 & 0.03  & 0   & 0.59 & 0.1   \\
\enddata
\end{deluxetable*}

We further calculate the cross-correlation of various combinations of hard and soft LCs and their derivatives within the flare (\#2). The results are presented in Table\,\ref{tab:crosscorr}. No other combination, but $dL_S/dt \propto L_H$ exhibits a significant correlation.
This is also the only correlation that is consistently strong in both XRISM and XMM-Newton and does not have a significant anti-correlation peak,
ruling out this correlation being a symptom of generic flare shapes.

\begin{table}[h]
\hspace{-1.5cm}
\small
\begin{tabular}{cc|ccc|ccc}
\hline\hline
\multicolumn{2}{c|}{} & \multicolumn{3}{c}{\textit{XRISM}/Xtend} & \multicolumn{3}{|c}{\textit{XMM-Newton} EPIC-PN} \\
\hline
\multicolumn{2}{c|}{Correlated series} & Strength & p-value & Lag (ks) & Strength & p-value & Lag (ks) \\
\hline
\textbf{$L_H$} & $\boldsymbol{\frac{d}{dt}}$\textbf{$L_S$} & \textbf{0.84} & \textbf{0.002} & \textbf{6} & \textbf{0.92} & \textbf{0.02} & \textbf{6} \\
$\frac{d}{dt}L_H$ & $L_S$               & 0.36 ($-$0.36) & 0.2 (0.2)    & 54 ($-$6)  & 0.44 ($-$0.45) & 0.2 (0.2)  & 48 ($-$6)  \\
$L_H$               & $\frac{d}{dt}L_H$ & 0.64 ($-$0.57) & 0.006 (0.02) & 6 ($-$6)   & 0.56 ($-$0.48) & 0.05 (0.1) & 6 ($-$6)   \\
$L_S$               & $\frac{d}{dt}L_S$ & 0.49 ($-$0.60) & 0.4 (0.3)    & 36 ($-$24) & 0.58 ($-$0.44) & 0.5 (0.4)  & 18 ($-$18) \\
$\frac{d}{dt}L_H$ & $\frac{d}{dt}L_S$ & 0.43           & 0.1          & 0          & 0.46 ($-$0.40) & 0.1 (0.2)  & 0 (12)     \\
\hline
\end{tabular}
\caption{Correlations between different combinations of hard ($L_H$) and soft ($L_S$) X-ray light curves and their derivatives during the Neupert flare. Correlations are calculated using XRISM/Xtend and XMM-Newton/EPIC-PN light curves at 6\,ks binning. The XRISM hard and soft X-rays are taken in the 7.2-12.0\,keV and the 0.4-2.0\,keV bands, respectively, while in XMM-Newton the 4.8-6.0\,keV band is used for hard X-rays. Where there is both a significant correlation and anti-correlation peak, the anti-correlation is presented in parentheses. The strongest correlation is found  between the hard X-ray LC and the derivative of the soft X-ray LC (highlighted), which is the solar-like Neupert effect.}
\label{tab:crosscorr}
\end{table}

\section{Time derivative calculation}

The time derivative of a light curve is computed with the \texttt{numpy} library in python using second order central differences in the interior points and first order one-sided differences in the boundaries. For interior points, the derivative can be approximated as

\begin{equation}
    \left(\frac{dy}{dt}\right)\approx \frac{y_{i+1}-y_{i-1}}{t_{i+1}-t_{i-1}}.
\end{equation}

Assuming uncorrelated measurement errors, uncertainties on the derivative are calculated using standard first order error propagation. At interior points the error of the derivative at each time $t_i$ is calculated as

\begin{equation}
    \sigma^2_{(dy/dt)_i} = \frac{\sigma^2_{(y)_{i+1}}+\sigma^2_{(y)_{i-1}}}{(t_{i+1}-t_{i-1})^2}
\end{equation}

\noindent where $\sigma_{(y)_i}$ is the error at $t_i$. At the boundaries uncertainty is calculated similarly using the relevant points for forward and backward one-sided differences.

\section{Cross Correlation}
\label{appendix_ccr}

To calculate the Cross Correlation Function (CCF) of two uniformly sampled time series, each series is standardized as $y \rightarrow (y-\bar{y})/\sigma_y$, where $\bar{y}$ and $\sigma_y$ are the mean and standard deviation of the measurements. The discrete CCF is calculated using the \texttt{scipy.signal.correlate} python function, and the corresponding lag indices are obtained with the \texttt{scipy.signal.correlation\_lags} function. The CCF is normalized by the length of the series to obtain the correlation strength.

To estimate the significance of the correlation, we calculated the p-value from the t-test statistic, which was calculated using the correlation strength $r$

\begin{equation}
    t=r\sqrt{\frac{N_{eff}-2}{1-r^2}}
\end{equation}

\noindent where $N_{eff}$ is the effective number of the degrees of freedom of the time series. Because each time series is not comprised of independent measurements, their lag-1 autocorrelation scores were used to calculate $N_{eff}$ as

\begin{equation}
    N_{eff}=N\frac{1-r_1 r_2}{1+r_1 r_2}
\end{equation}

\noindent where $N$ is the number of measurements and $r_1$, $r_2$ are the autocorrelation strengths of the two time series \citep{Bretherton1999}.

%\begin{contribution}
%%This section gives authors the space to recognize author contributions. The text inside this environment is NOT counted towards the total word quanta. At a minimum, manuscripts are expected to include this text:

%[what should we put here?]

%% But authors are expected to provide more specific details, e.g. 
%%
%%SC was responsible for writing and submitting the manuscript.
%%WWM came up with the initial research concept and edited the manuscript.
%%OTS obtained the funding and edited the manuscript.
%%EBF provided the formal analysis and validation. He also edited the manuscript.
%%GEH Supervised the undergraduates, wrote the software and administers the project github and Zenodo repositories.
%%
%% Authors can use the Contributor Role Taxonomy (CRediT) at
%% for ideas on how write a good statement tailored to their needs.

%\end{contribution}

\facilities{XRISM (Xtend), XMM-Newton (OM and EPIC-PN)}

\software{numpy \citep{numpy}, 
          scipy \citep{scipy},
          HEASoft \citep{Heasoft},
          Xspec \citep{xspec1996},
          XMM-Newton SAS \citep{SAS2004},
          }

\bibliography{references}{}
\bibliographystyle{aasjournalv7}



\end{document}